\documentclass[aps,pra,preprint,superscriptaddress]{revtex4-1}
\usepackage{amsmath}
\usepackage{graphicx}
\usepackage{subfigure}
\usepackage[colorlinks=True,linkcolor=blue,anchorcolor=blue,citecolor=blue,CJKbookmarks=true]{hyperref}

\usepackage{ulem}

\begin{document}

\title{Effects of control fields on the pair creation and the vacuum information transmission}

\author{J. X. Wu}
\affiliation{School of Science, China University of Mining and Technology, Beijing 100083, China}
\author{C. Gong}
\affiliation{Department of Mathematics and Physics, North China Electric Power University, Baoding 071003, China}
\author{A. R. Sun}
\affiliation{School of Science, China University of Mining and Technology, Beijing 100083, China}
\author{Z. L. Li}
\email{Corresponding author: zlli@cumtb.edu.cn}
\affiliation{School of Science, China University of Mining and Technology, Beijing 100083, China}
\author{Y. J. Li}
\email{Corresponding author: lyj@aphy.iphy.ac.cn}
\affiliation{School of Science, China University of Mining and Technology, Beijing 100083, China}
\affiliation{
  State Key Laboratory for GeoMechanics and Deep Underground Engineering, \\
  China University of Mining and Technology, Beijing 100083, China}

\date{\today}

\begin{abstract}
The effects of control fields on the energy spectra and the number of created pairs and the information transmission by the Dirac vacuum modes are investigated by employing computational quantum field theory approach.
It is found that the oscillation structures on the energy spectra are sensitive to the direction, the width, and the oscillation frequency of control fields. The pair yield can have obvious changes for a small frequency and a very large frequency.
Moreover, the information encoded in the control fields, such as the field direction, the laser frequency and the time interval between two laser pulses, can also embodied by the vacuum modes in the change of pair-creation rate with time.
These results not only can deepen our understanding of the control of pair creation and the information transmission, but also can provide a theoretical reference to the related experiments in the future.
\end{abstract}

\maketitle

\section{Introduction}

Since the positrons were put forward theoretically by Dirac \cite{Dirac1928} and discovered experimentally by Anderson \cite{Anderson1933}, people realized that
vacuum is not stable, and will decay and spontaneously create electron-positron pairs in extreme situations \cite{Greiner1985}.
Therefore, the study of vacuum decay and electron-positron pair creation has attracted a large number of scholars to conduct related research \cite{Piazza2012, Xie2017}.
In 1951, Schwinger calculated the vacuum decay rate in a static strong electric field \cite{Schwinger1951}, i.e., ($\varGamma \sim \exp(-\pi E_{\rm cr}/E)$), and the critical field strength $E_{\rm cr} $ of vacuum decay is $1.32\times 10^{16}\ {\rm V/cm} $.
However, due to the requirement of electric field strength is huge, the current electric field intensity or laser intensity is far away from reaching the theoretical intensity, the electron-positron pair creation from the vacuum has not been directly observed experimentally.
This is the joint effort of theoretical physicists and experimental physicists, either reducing the laser requirements theoretically, or continuing to increase the laser intensity experimentally.
With the rapid development of laser technology \cite{laser}, it is expected that the laser can be directly used to break down the vacuum and create electron-positron pairs.

Vacuum pair creation in strong fields can be affected by many factors, such as temporal interference effects \cite{Akkermans2012}, and significant differences caused by the spatial inhomogeneity of the electromagnetic field \cite{lvdongli2018}.
In 2018, Lv et al. \cite{Lv2018} studied the effect of a subcritical electric field which is placed far outside the pair creation zone on the energy spectrum of positrons created by the localized supercritical field.
They found a counterintuitive phenomenon that the spatially localized subcritical field can also induce the obvious oscillation of the energy spectra of created positrons which have never been to the subcritical field.
In general, however, the spatially localized subcritical field which is also called the control field varies with the time, for instance, the laser field. So what is the effect of the time-dependent control field on the oscillation structure of the energy spectra of created particles?
This is one of our researches including the effect of the width of control fields on the energy spectra.
Furthermore, in 2019 Su et al. \cite{Su2019} proposed that Dirac vacuum modes could be used as a potentially loss-free carrier of information transmission.
Specifically, the temporal shape of the control field is encoded with a binary sequence of distinguishable high and low values of the amplitude, and this information can be embodied by the vacuum mode in the pair-creation rate corresponding to the supercritical field for a selected energy. Except for the binary encoding of the control field, is it possible to encode the control field with laser parameters, such as the frequency and the time interval between two laser pulses, and transport these information by the vacuum modes? This is another researches of this work.
In addition to the above discussions, the vacuum can also show the Casimir effect \cite{Chan2018, Rodriguez-Lopez2017}, vacuum birefringence effect \cite{Capparelli2017} and photon-photon scattering \cite{King2010}, etc., so it has an important implication for exploring the structure of the vacuum.

This paper is organized as follows. In Sec. \ref{sectwo} we introduced the theoretical framework-
computational quantum field theory(CQFT) \cite{Braun1999} and the model style we explored.
In Sec. \ref{secthree} we studied the effect of direction, width, and time-dependent of the localized control field on pairs created by supercritical fields.
In Sec. \ref{secfour} the pair-creation process can receive some information encoded in the control field and act as a medium for information transmission.
A summary and discussion are give in Sec. \ref{secfive}. Appendix shows the analytical solution of the transmission coefficient for a double step potential.

\section{Theoretical approach and external field model \label{sectwo}}
The starting point is the Dirac equation
$i\hbar\frac{\partial \psi}{\partial t}=H\psi$,
where $H=c\pmb{\alpha\cdot}\mathbf{p}+mc^2\beta+eV(\mathbf{x},t)$ is the Hamiltonian, $\pmb{\alpha}=(\alpha_1,\alpha_2,\alpha_3) $ and $ \beta $ denote $4\times4$ Dirac matrices, $\mathbf{p}=-i\hbar\pmb{\nabla}$ is the canonical momentum operator, $c$ is the speed of light, $e$ and $m$ are the particle charge and mass, $V(\mathbf{x}, t)$ is the scalar potential.
For convenience, the atomic units, $\hbar = m =e= 1$ and  $c=137.036$, are used in the following content.
In the one-dimensional case, the Hamiltonian can be simplified as
\begin{equation}
  H=c\sigma_{1\ }p_x+c^2\sigma_3+V(x,t),
\end{equation}
where $\sigma_1$ and $\sigma_3$ are Pauli matrices.

In the framework of quantum field theory (QFT), the Dirac field operator can be expanded with different complete basis as
\begin{equation}
  \begin{matrix}
    \hat \Psi(x,t)&=\sum_{p}{\hat{b}_p(t)u_p(x)}+\sum_{n}{\hat{d}_n^\dag(t)v_n(x)}\\&
    =\sum_{p}{\hat{b}_pu_p(x,t)}+\sum_{n}{\hat{d}_n^\dag v_n(x,t)}\\
  \end{matrix}
\end{equation}
where $u_{p}(x)$, $v_{n}(x)$ are the positive and negative solutions to free particles, and  $u_{p}(x,t)$, $v_{n}(x,t)$ are the solutions to particles in external fields which can be obtained by evolving the $u_{p}(x)$and $v_n{x}$ according to Dirac equation. $\hat b_p^\dag$ and $\hat b_p$ are the creation and annihilation operators of electrons,
and $\hat d_n^\dag$ and $\hat d_n$ are the creation and annihilation operators of positrons. Using
the anticommutation relationship between the creation operator and the
annihilation operator, one can get
\begin{equation}
  \begin{split}
    \hat b_p(t)=\sum_{p^\prime}{\hat b_{p^\prime}U_{pp^\prime}(t)}+\sum_{n}{\hat d_n^\dag U_{pn}(t)}\\
    \hat d_n^\dag(t)=\sum_{p}{\hat b_pU_{np}(t)}+\sum_{n'}{\hat d_{n'}^\dag U_{nn^\prime}(t)}
  \end{split}
\end{equation}
where $U_{pp^\prime}(t)=\int dxu_p^\dag(x)u_{p^\prime}(x,t)$, $U_{pn}(t)=\int dxu_p^\dag(x)v_n(x,t) $, $U_{np}(t)=\int dxv_n^\dag(x)u_p(x,t)$, and $U_{nn^\prime}(t)=\int dxv_n^\dag(x)v_{n^\prime}(x,t)$.
Similarly, using the conjugate form of $\hat \Psi(x,t)$, the expression of $\hat b_p^\dag(t)$ and $\hat d_n(t)$ can be obtained.
The probability density of created electrons is defined as
\begin{equation}
  \begin{matrix}
    \rho(x,t)&=\langle \mathrm{vac}|\hat{\Psi}_e^\dag(x,t)\hat{\Psi}_e(x,t)|\mathrm{vac}\rangle\\
    &=\sum_{n}{\sum_{p}{{|U}_{pn}(t)u_p(x)|}}^2\\
  \end{matrix}
\end{equation}
where $\hat{\Psi}_e(x,t)=\Sigma_p\hat{b}_p(t)u_p(x)$ is the electron portion of the field operator.
Therefore, the particle yield $N(t)$ can be obtained from the integration of the particle probability density over space, i.e., $N(t) = \int dx \rho(x,t)$.
Finally, we have
\begin{equation}\label{nt}
  N(t)=\Sigma_{p,n}{|U_{pn}(t)|}^2,
\end{equation}
where $U_{pn}(t)$ is also the projection of the time-evolving negative energy state on the free positive energy state.
The evolution of the negative energy state can be obtained by using the split-operator method \cite{Braun1999, Mocken2008}.
From Eq. (\ref{nt}), one can also obtain the number of created electrons with the momentum $p$, i.e., $N_p(t)=\Sigma_n{|U_{pn}(t)|}^2$, which is a series of discrete points on the momentum distribution function of created particles.
Since the created pair yield can also be expressed as $N(t)=\int dE N(E,t)$ by the energy spectrum $N(E,t)$, it is easy to find that $N(t)=\Sigma_{E}\Delta E N(E, t)=\Sigma_{p} (dE/dp) \Delta p N(E(p), t)=\Sigma_{p}N_p(t)$ in the box normalization, where $\Delta p=2\pi/L$ and $L$ is the box length.
Then the energy spectrum of created particles
\begin{equation}\label{energyspectrum}
N(E, t)=\frac{L}{2\pi(dE/dp)} N_p(t)=\frac{L}{2\pi(dE/dp)}\Sigma_n{|U_{pn}(t)|}^2,
\end{equation}
where $dE/dp$ is the group velocity of the particle wavepacket.

In our model system, the scalar potential $V(x)=V_1S(x)+V_2S(x+d)$ is employed, which $S\left(x\right)=[1+\tanh(x/w)]/2$ is the Sauter potential, $w$ is the electric field width.
Here we choose the potential height of the control field $V_2<2c^2$, and the height of the supercritical field $V_1>2c^2$.
The corresponding electric field is
\begin{eqnarray}
  \mathcal{E}(x)=&&-V_1/\left(2w\right)\{1-\left[\tanh{\left(x/w\right)}\right]^2\}\nonumber\\
  &&-V_2/\left(2w\right)\{1-\left[\tanh{\left(\left(x+d\right)/w\right)}\right]^2\}.
  \label{ele}
\end{eqnarray}

\begin{figure}
\includegraphics[width=0.45\textwidth]{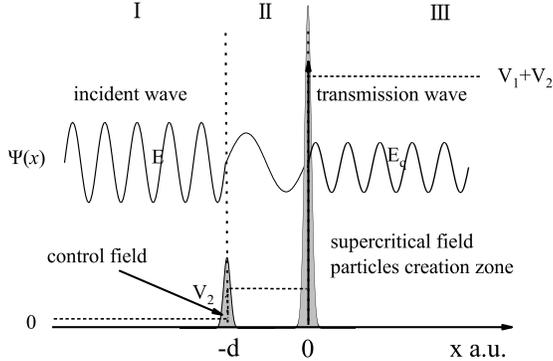}
\caption{A sketch of our model, a localised supercritical field ($x=0$) and a control field ($x=-d$) are placed, and their potential distributions are shown in the diagram. We also show roughly the quantum mechanical scattering process in the potential barrier. \label{model}}
\end{figure}

The sketch of this potential and the corresponding electric field are shown in Fig. \ref{model}.
Note that there is no overlap in space between the control field located at $x=-d$ and the supercritical field located at $x=0$.
However, the energy spectrum of created particles can be affected by the control field outside the pair creation zone.

\section{numerical result\label{secthree}}
\subsection{Effect of the direction of control fields on pair creation}
\begin{figure*}
\subfigure{
\includegraphics[width=0.45\textwidth]{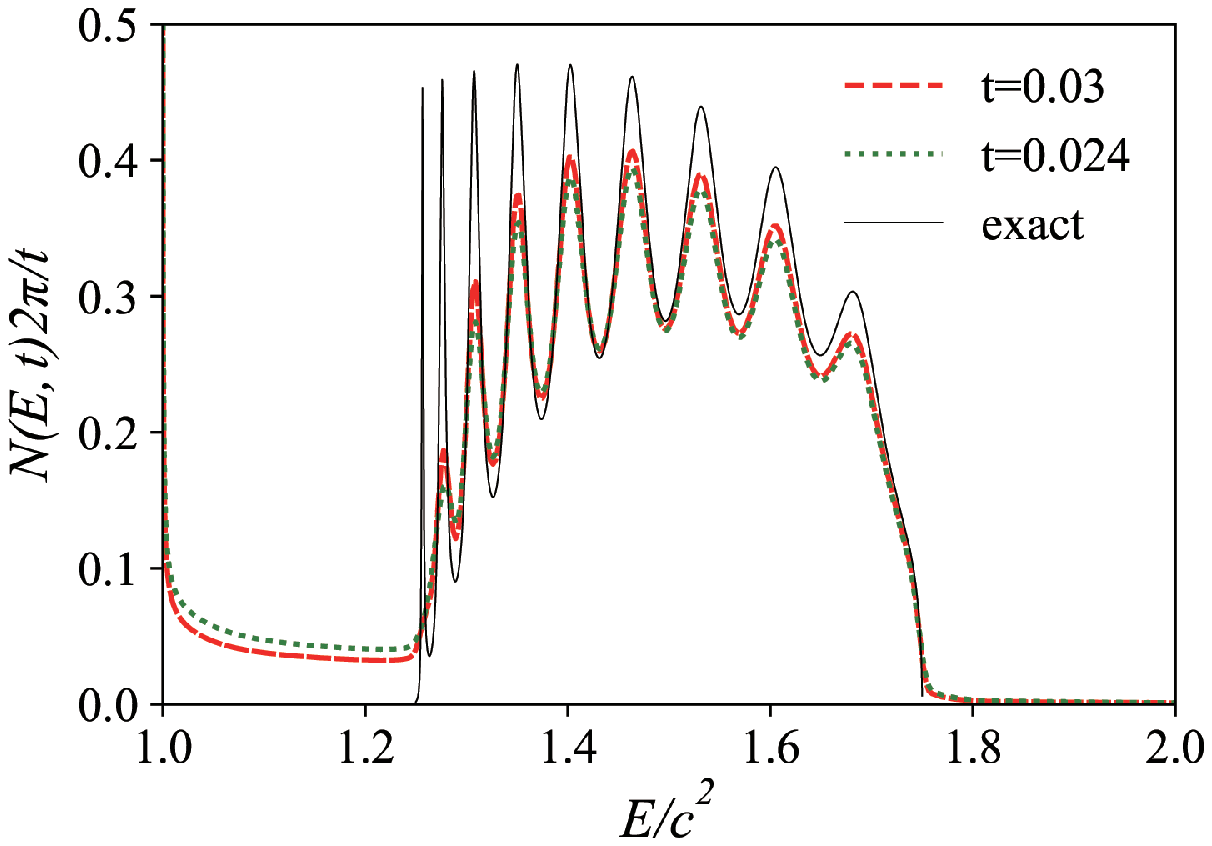}
}
\subfigure{
\includegraphics[width=0.45\textwidth]{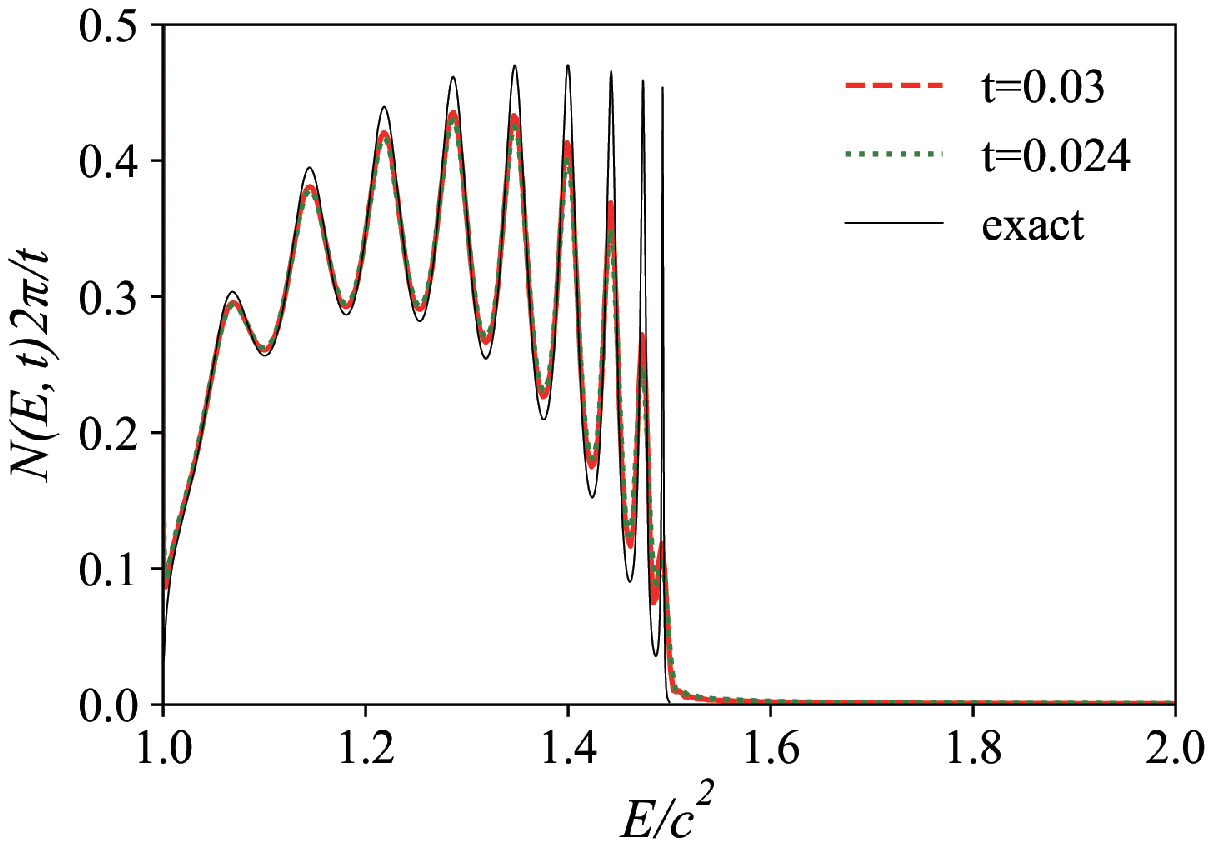}
}
\caption{Energy spectra of positrons (left) and electrons (right), the black line is the transmission coefficient of double-step barrier according to Eq. (\ref{tf}), the red dashed line and the green dotted line are our numerical results, the red dashed line is the (normalized) energy spectra $N(E,t)2\pi/t$ with the system evolution $t=0.03$a.u., and the green dotted line one is $t=0.024$a.u., the other parameters are $V_1=2.5c^2, V_2=0.25c^2, L=12.0{\rm a.u.}, d=0.2{\rm a.u.}, w=0.075/c, N_{\rm t}=6000$ and $N_{\rm x}=8192$.
\label{engyspe}
}
\end{figure*}

We consider two types of electric field. One is that the direction of the control field is the same as the direction of the supercritical field.
The other is that the direction of the control field is opposite to the direction of the supercritical field.
The former one can form a double-step potential, see Fig. \ref{model}.
For this potential, the energy spectra of created electrons and positrons can be obtained by calculating the transmission coefficient of electrons, see Appendix for details.
\begin{equation}
  T_E=\frac{4\gamma\tau}{{(\gamma\tau+1)}^2-(\gamma^2-1)(\tau^2-1){\sin}^2(kd)}
  \label{tf}
\end{equation}
where $\gamma=\sqrt{\frac{E-c^2}{E+c^2}\frac{E-V_1+c^2}{E-V_1-c^2}}$, $\tau=\sqrt{\frac{V_1+V_2-E+c^2}{V_1+V_2-E-c^2}\frac{E-V_1-c^2}{E-V_1+c^2}}$, and $k=\sqrt{{-c}^4{+\left(E-V_2\right)}^2}/c$.
In the Klein range $c^2<E<V_1-c^2 $, this analytical results (solid black line) are well agree with our numerical results (dashed red line) obtained by Eq. (\ref{energyspectrum}), see Fig. \ref{engyspe}. The difference between these two results can be further reduced with the increase of time.
This is because the analytical results are given at infinity time, whereas the numerical results only show the energy spectra in a finite time.

The energy spectra of created particles can be understood from the point of view of Dirac's hole theory.
For the supercritical field, there is a cross between the positive energy continuum and the negative one in the energy range of $1.25c^2$ to $1.75c^2$.
Therefore, the particles in the Dirac sea with an energy between $1.25c^2$ and $1.75c^2$ can tunnel into the positive energy continuum and become real particles, meanwhile the holes left in the Dirac sea are interpreted as the antiparticles.
However, due to the existence of the control field, the tunneling particles should also go through a barrier with height $0.25c^2$, which causes the resonance transmission and leads to the oscillation of the energy spectra.
Another interesting phenomenon is that the electron never reaches the control field, but its energy spectrum has and oscillatory structure as well, which is completely incomprehensible in classical physics. The reason for this is that the resonance transmission will enhance the creation of particles with the resonance energy.

\begin{figure*}
  \subfigure{
    \includegraphics[width=0.45\textwidth]{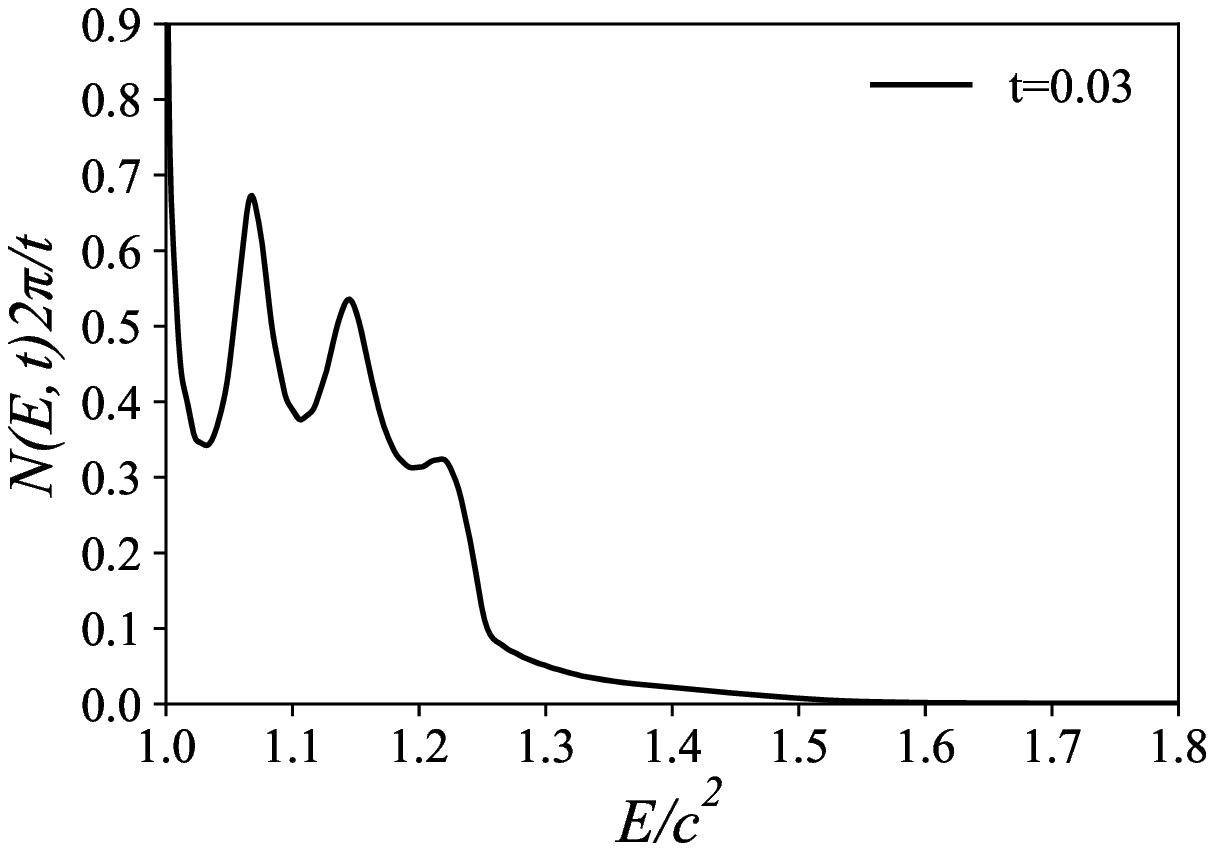}
  }
  \subfigure{
    \includegraphics[width=0.45\textwidth]{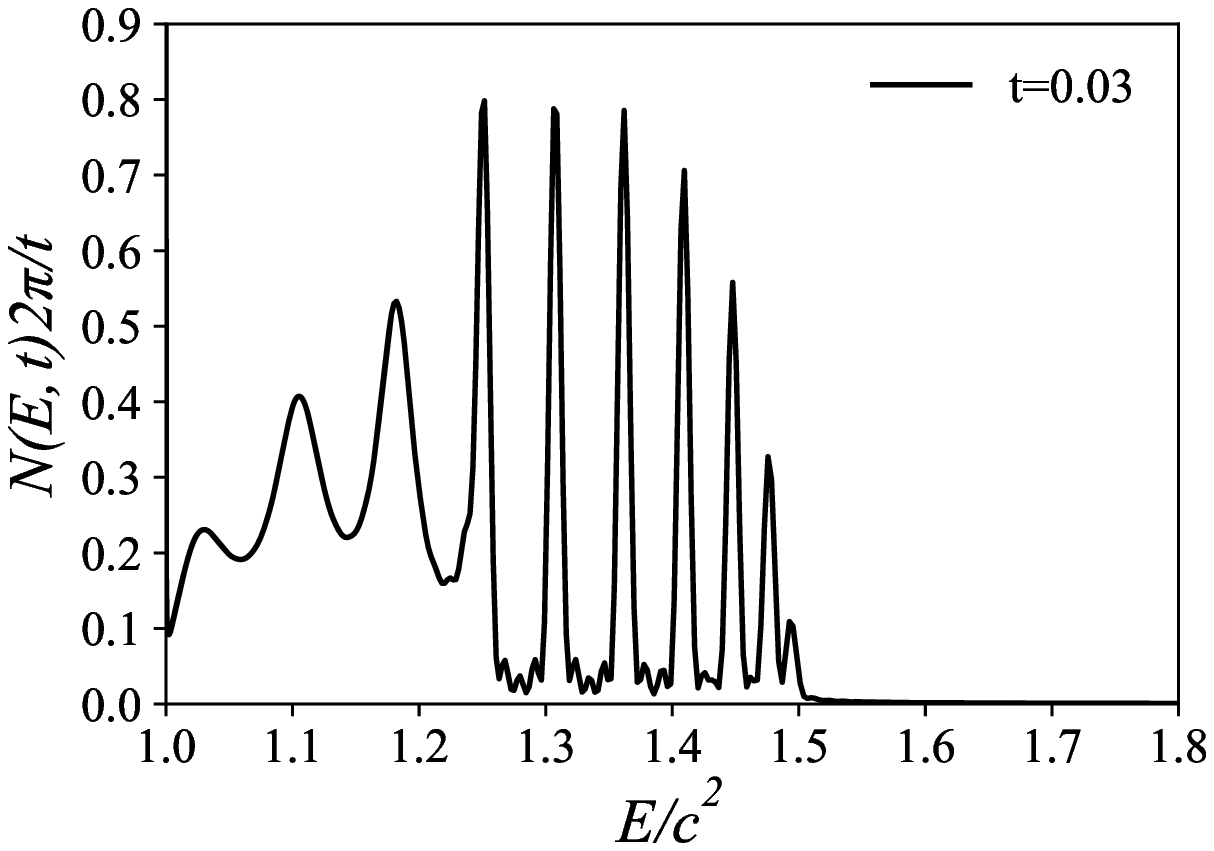}
  }
  \caption{Energy spectra of positrons (left) and electrons (right) in the asymmetric potential well. Here $V_1=2.5c^2$, $V_2=-0.25c^2$, the other parameters are the same as Fig. \ref{engyspe}. \label{rev_engy_spe}}
\end{figure*}

When the direction of the control field is opposite to the direction of the supercritical field, an asymmetric potential well is formed, and the energy spectra of created particles in this field are calculated numerically and shown in Fig. \ref{rev_engy_spe}.
One can see that although the energy spectra still have obvious oscillatory structures, there are two differences from the results in the case of double-step potential.
One is the presence of discrete peaks on the energy spectrum of electrons.
The other is the symmetry breaking between the energy spectra of positrons and electrons.
These can be understood from the perspective of quantum mechanical scattering picture.
When the energy of the incident wave is in the range $c^2-|V_2|<E<c^2$, the momentum in region I of the potential is imaginary, but in region II it is real.
This indicates that there are discrete bound states in the asymmetric potential well, which can also induce the pair creation, but only the particles with the energy of bound states can be produced.
According to the formula of the bound state levels in a symmetric potential well \cite{Greiner1997}, $ck\cot(kd)=-(EV_2)/(c\kappa)-c\kappa$ (where $E$ and $\kappa$ are energy and the imaginary part of momentum in region I), one can estimate that there are approximately seven bound state levels in our asymmetric potential well. This just corresponds to the discrete peaks of the energy spectrum of electrons.
Additionally, since the created low-energy positrons cannot pass through the potential barrier formed by the control field and gather in the asymmetric potential well, the number of created positrons with the energy between $c^2$ and $1.25c^2$ is higher than that of created electrons. This also causes the symmetry breaking between the energy spectra of positrons and electrons.

\begin{figure}
  \includegraphics[width=0.4\textwidth]{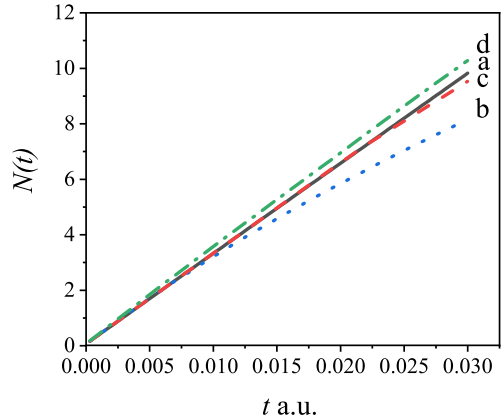}
  \caption{Pair yield evolving with time for different control fields. Curve a (solid black line) corresponds to the pair yield only in a supercritical field, which is also the result that the direction of control field is the same as the supercritical field. Curve b (dotted blue line) is the result that the direction of the control field is opposite to the supercritical field. Curve c (dashed red line) is the result that the control field varies sinusoidally with time and the time of evolution is just a period ($\omega=0.011c^2$). Curve d (dot-dashed green line) is the result that the control field varies sinusoidally with time and $\omega=2c^2$.
  The other parameters are the same as Fig. \ref{engyspe}. \label{N_t_diff_diret}}
\end{figure}

In order to study the effect of control fields on the pair yield, we plot the number of created pairs as a function of time, see Fig. \ref{N_t_diff_diret}.
One can see that when the direction of the control field is the same as the supercritical field, the presence of control field has no effect on the total number of particles created by the supercritical field except inducing the oscillation of energy spectra.
However, when the direction of the control field is opposite to the direction of the supercritical field, the pair-creation rate is gradually decreasing with time.
The reason is that the created particles move to the left and enter the control field, and then rebound to the supercritical region by the control field, which suppresses the particle pair yield \cite{Krekora2004}.

\subsection{Effect of the width of control fields on pair creation}
\begin{figure*}[!ht]
  \subfigure{
    \includegraphics[width=0.45\textwidth]{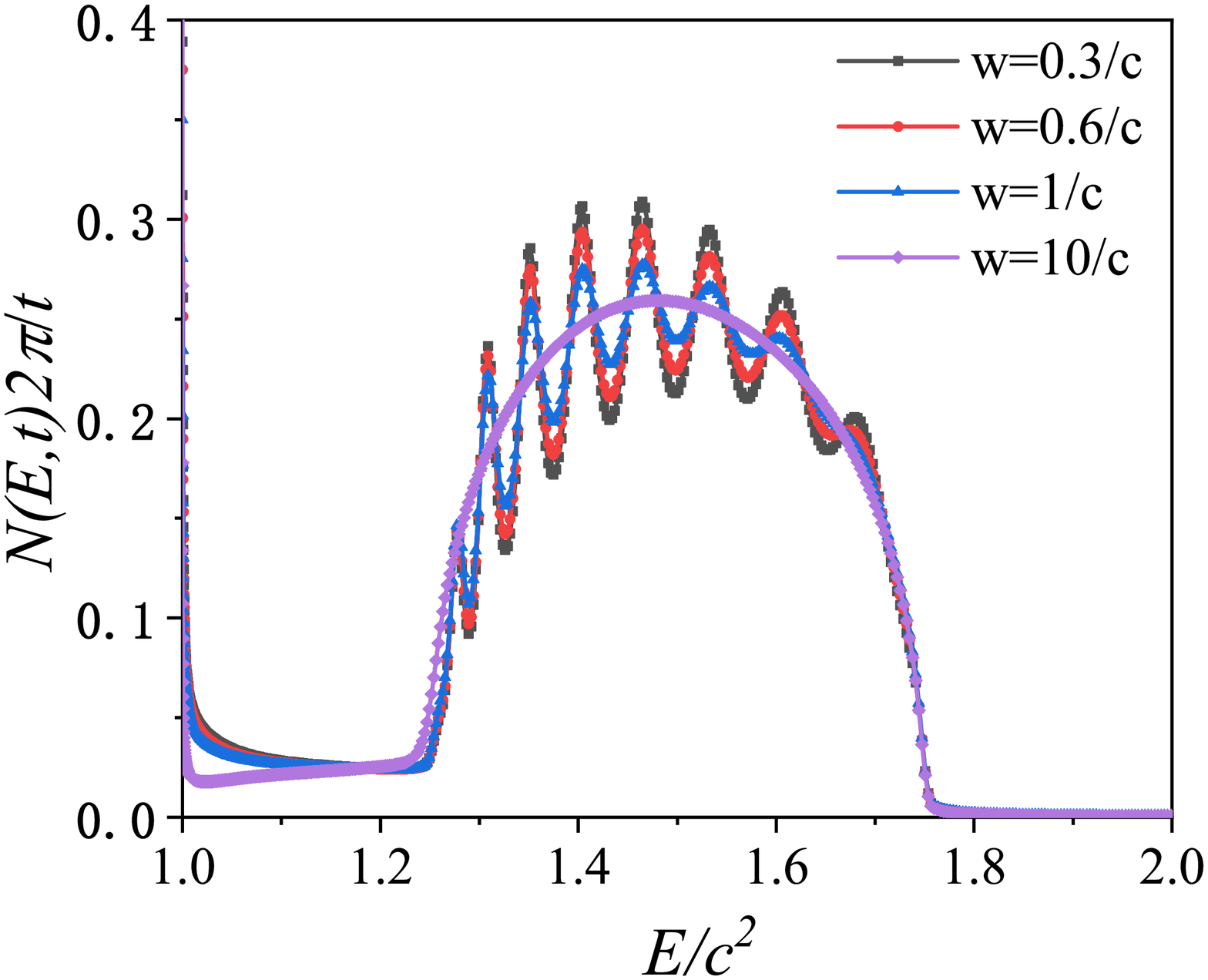}
  }
  \raisebox{.13cm}{\subfigure{
    \includegraphics[width=0.45\textwidth]{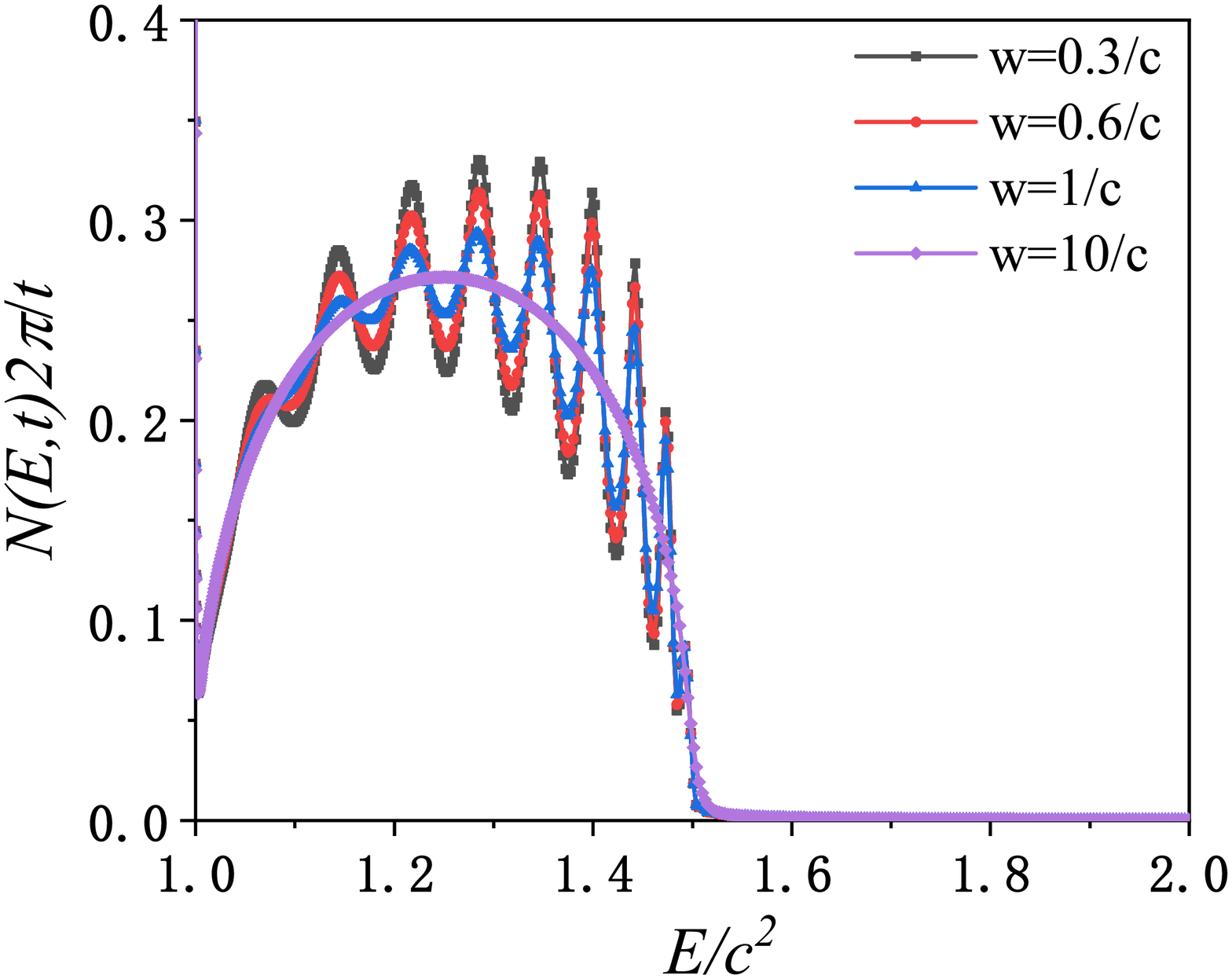}
  }}
  \caption{Energy spectra of created positron (left) and electron (right) for different widths of control fields.
  The field parameters are $V_1=2.5c^2,L=12.0{\rm a.u.},V_2=0.25c^2,T_{\rm max}=0.03{\rm a.u.}, N_{\rm x}=8192, N_{\rm t}=6000$.\label{width} }
\end{figure*}

In this subsection we will discuss the effect of the width of control fields on the pair creation process.
Note that the width of the supercritical field is kept constant but the width of the control field is variable.
The energy spectra of created particles for the width of the control field $w=0.3/c,0.6/c,1/c,10/c$ are shown in Fig. \ref{width}. It is found that the width of the control field does not affect the created pair yield, but it will change the final energy distribution of the created pairs.

Moreover, with the increase of the width of the control field, the oscillation of the energy spectra becomes less obvious.
When the width is 10/c, the energy spectra become smooth and similar to the energy spectra of created particles in a single supercritical field.
This is also understandable from the perspective of quantum
mechanical scattering picture.
In the over-the-barrier regime for which the incoming energy of particles is larger than the potential height formed by the control field, the transmission coefficient displays obvious resonance transmission phenomenon for the zero field width.
However, with the increase of the field width, the position of the resonance transmission does not change, but the oscillation of transmission coefficient becomes less obvious \cite{Kennedy_2002}. From another perspective, the increase of the width of the control field corresponds to the decrease of the electric field strength, which will continually reduce the effect of control field on the pair creation by the supercritical field.

\subsection{Effect of time-dependent control fields on pair creation}
Here the control field we used has the form
\begin{equation}\label{V2xt}
V_2\left(x,t\right)=V_2S(x+d)\sin(\omega t+\varphi _0),
\end{equation}
where $S(x)$ is the Sauter potential, $\omega$ is the oscillation frequency, and $\varphi_0$ is the phase.
In this situation, the control field will affect all aspects of particles created by the supercritical field.

\begin{figure*}[!ht]
  \subfigure{
    \includegraphics[width=0.45\textwidth]{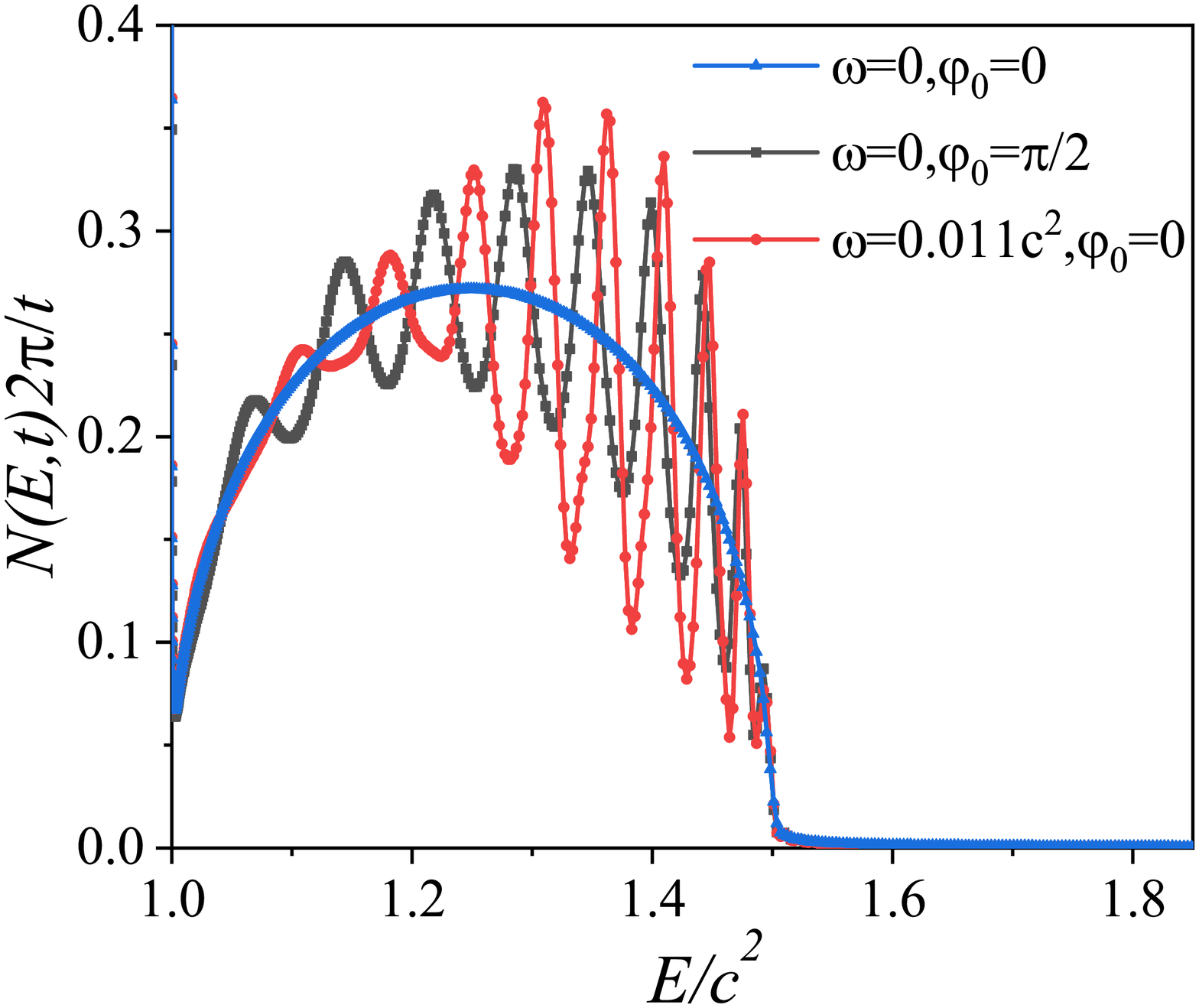}
  }
   \raisebox{.1cm}{\subfigure{
    \includegraphics[width=0.45\textwidth]{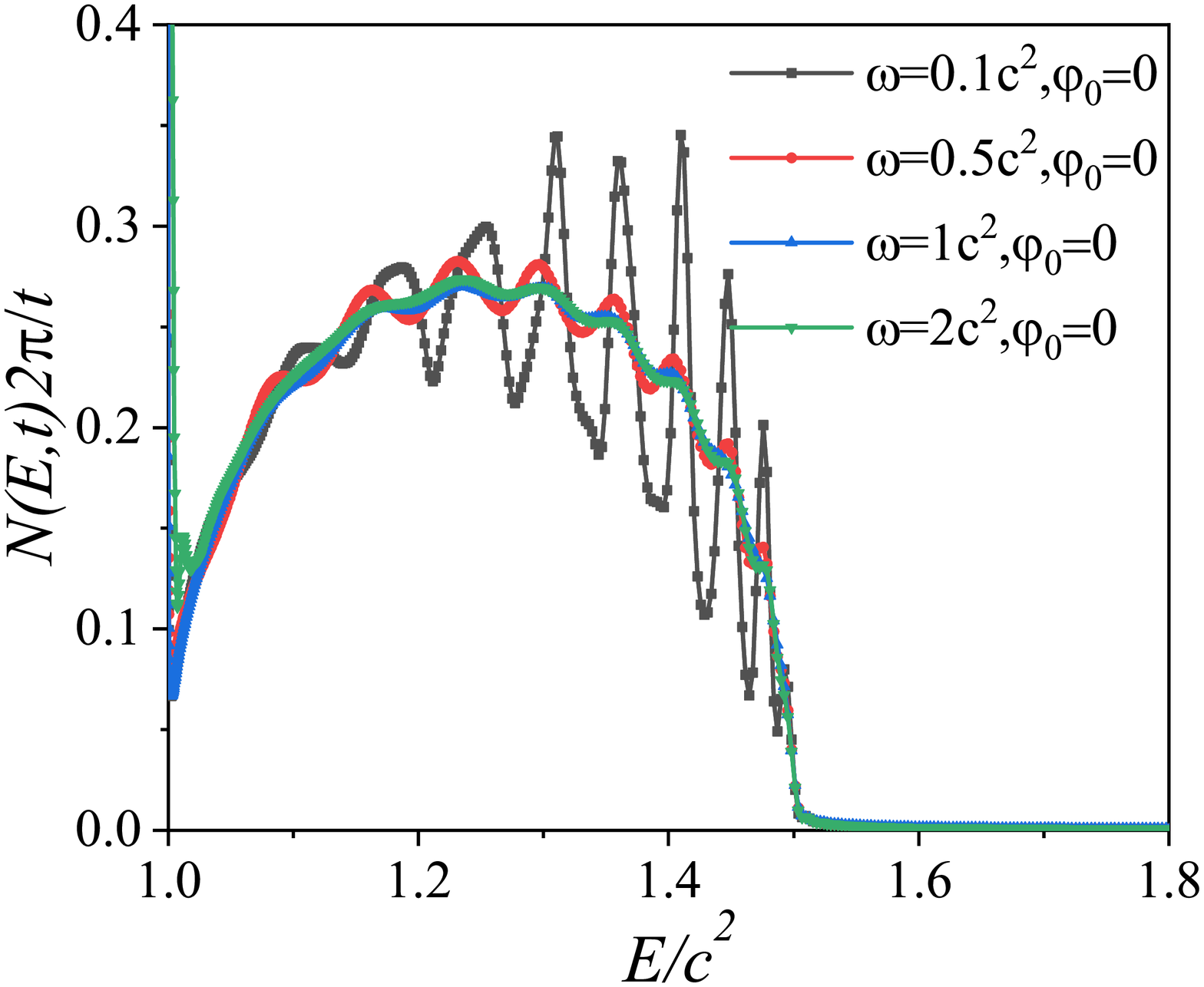}
  }}
  \caption{Electron energy spectra for different frequencies of the control fields.
	To make the results clear, we split the results into two figures, the left one $\omega=0, \varphi _0=\pi /2$, $\omega=0.011c^2,\varphi _0=0$, and the right one
  $\omega=0.1c^2,0.5c^2,1c^2,2c^2, \varphi _0=0$, the other parameters are
  $V_1=2.5c^2,V_2=0.25c^2,L=12.0{\rm a.u.},T_{\rm max}=0.03{\rm a.u.}, N_{\rm x}=8192, N_{\rm t}=6000$.
   \label{timespec}
  }
\end{figure*}

First, we calculate the energy spectra of created pairs for the frequency of control fields $\omega=0.011{c}^2, 0.1{c}^2, 0.5{c}^2, 1{c}^2, 2{c}^2$, see Fig. \ref{timespec}.
Note that for $\omega=0$ and $\varphi_0=0$ the control field vanishes and for $\omega=0$ and $\varphi_0=\pi/2$ the control field has the same direction as the supercritical field.
One can see that the energy spectra of electrons are clearly disturbed by the frequency of the control field.
For a small frequency, such as $\omega=0.011c^2$ or $0.1c^2$, the location of the peaks of energy spectra is drifted and the amplitude of the oscillation becomes greater.
However, with the increase of the frequency, the oscillation becomes less obvious. The reason is that this oscillation is averaged by the rapid change of the direction of control fields.

The effect of time-dependent control fields on the the number of created pairs are shown in Fig. \ref{N_t_diff_diret}. It can be seen that for $\omega=0.011{c}^2 $ (curve c) the particle creation rate starts to become small after $t=0.015$ a.u..
This is because that the time-varying control field just has a cycle during the evolution time ($T_{\rm max}=0.03$ a.u.) and the direction of the control field changes at $t=0.015$ a.u..
Based on our previous discussion, when the direction of the control field is opposite to the direction of the supercritical field, the pair yield becomes small.
Therefore, when the frequency is small, the pair creation will inevitably be suppressed.
But as the control field frequency increases, the probability of the created positron being bounced back into the creation region by the control field also becomes smaller.
Thus the suppression effect is not obvious.
Continuing to increase the frequency until the control field can also create substantial pairs by multiphoton absorption, the pair-creation rate become large again.
When $\omega=2c^2$, the pair-creation rate is significantly greater than that for $\omega<2c^2$, that is because the Dirac sea can directly absorb photons from the control field and create electron-positron pairs.

\section{pair creation and information transmission\label{secfour}}
Based on the previous discussions, one can see that the presence of a control field which is subcritical and far away from the supercritical field can still affect the pair creation by the supercritical field.
So it is possible to encode the control field with different temporal information and probe whether the information is embodied in the change of the pair-creation rate with time.

Since the pair-creation rate of created electrons with specific energy is sensitive to the change of the amplitude and direction of the control field, we focus on the change of pair creation rate of created electrons with energy $1.25c^2$. In this case, the pair creation will be suppressed when the control field and the supercritical field have the same direction, but it will be enhanced when these two fields have the opposite directions. Thus, the temporal information encoded in the control field can be embodied in the change of pair-creation rate.
Here we consider four types of control fields, i.e., $V_2(x,t)=V_2S(x+d)f(t)$, where
$$f(t)=
\begin{cases}
  -\sin(\omega t), & \\
  -\exp[-\frac{(t-0.015)^2}{2\sigma_1^2}]\sin(\omega t), &  \\
  -\exp[-\frac{(t-0.01)^2}{2\sigma_2^2}]-\exp[-\frac{(t-0.02)^2}{2\sigma_2^2}], & \\
  -\exp[-\frac{(t-0.006)^2}{2\sigma_3^2}]+\exp[-\frac{(t-0.024)^2}{2\sigma_3^2}], &
\end{cases} $$
and $\omega=0.1c^2$, $\sigma_1=0.005$a.u., $\sigma_2=0.001$a.u., $\sigma_3=0.002$a.u..
Note that the negative sign here means that the control field and supercritical field have the opposite directions initially.

\begin{figure*}
  \includegraphics[width=0.9\textwidth]{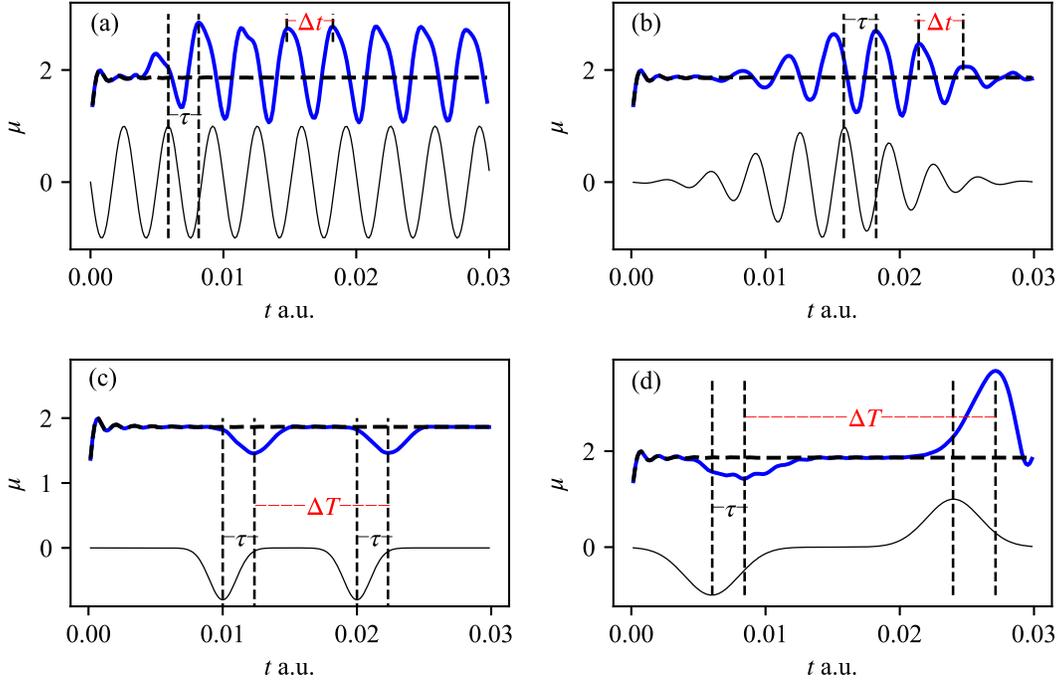}
  \caption{Pair-creation rate as a function of time for four types of the time-dependent control fields, see the solid blue line.
  The thin solid black line shows the time-dependent factor of the control field $f(t)$.
  The thick dashed black line denotes the pair-creation rate in the absence of the control fields. $\tau, \Delta t$, and $\Delta T$ represent the response time, period time, and time interval of two Gaussian envelope fields, respectively.  Other parameters are $V_1=2.5c^2,
  V_2=0.25c^2,T_{\rm max}=0.03{\rm a.u.},N_{\rm x}=8192,N_{\rm t}=6000, d=0.2{\rm a.u.}$. \label{mu-t}}
\end{figure*}

The pair-creation rate $\mu(t)$ and the time-dependent factor of control fields $f(t)$ are shown in Fig. \ref{mu-t}.
First, one can see that the pair-creation rate has an oscillation in the beginning.
This is caused by turning on the supercritical field rapidly and can be removed by turning on this field smoothly.
Second, the change of pair-creation rate has a delay time comparing with that of control fields.
By theoretical analysis, this response time corresponds to the transmission time of the right-traveling Dirac state moving from the control field to the supercritical field, which can be obtained by
\begin{equation}
  \tau=\frac{d}{v}=d\sqrt{c^4+c^2p^2}/cp
\end{equation}
where $d$ denotes the distance between the two electric fields, $v$ and $p$ are the group velocity and momentum of the right-traveling Dirac state, respectively.
For our case, the response time $\tau$ calculated by the above equation is $2.432\times 10^{-3}$, and that obtained by the numerical results is $2.285\times10^{-3},\, 2.412\times10^{-3},\,2.361\times10^{-3},\,2.412\times10^{-3}$, respectively.
It indicates that the above theoretical formula can be well fitted to our results within an acceptable error range.
After considering the response time, we find that the change of pair-creation rate with time can well express the temporal information of the control field.

Specifically, from Fig. \ref{mu-t}(a) and Fig. \ref{mu-t}(b), we find that the period of the oscillation of the pair-creation rate $\Delta t$ are $3.467\times10^{-3}$a.u. and $3.317\times10^{-3}$a.u., respectively.
The period of the control field with $\omega=0.1c^2$ is $2\pi/\omega=3.346\times10^{-3}$a.u..
It means that the frequency information encoded in the control field can be embodied in the change of the pair-creation rate with time.
Fig. \ref{mu-t}(c) and \ref{mu-t}(d) show the results for the last two forms of $f(t)$, i.e., $f(t)$ consisted of two identical Gaussian envelope fields and the two alternating ones.
For the former figure, the time interval $\Delta T$ between the two peaks of pair-creation rate is $0.009948$a.u.. For the latter one, it is $0.01869$a.u..
These results match the time interval of the two Gaussian envelope fields well: $ 0.01$ for identical case and $0.018$ for alternating case.
Therefore, the information of the time interval can also be well expressed by the change of pair-creation rate with time.

Based on the above discussions, we can see that electron-positron pair creation process can well receive the temporal information encoded in the control fields.
This is a novel kind of information transmission method without any material as a transport medium.
In the future, when the laser intensity reaches sufficient strength, the control fields may play an important role on the information transmission.

\section{Summary and Outlook\label{secfive}}
In summary, we investigate the effects of the profile of control fields on the energy spectra and number of created particles and vacuum information transmission.
First, it is found that the energy spectra and the number of particles created by the supercritical field is sensitive to the direction of the control field.
Particularly, the pair creation is suppressed when the direction of the control field is opposite to that of the supercritical field.
Second, the width of the control field also affects the pair creation.
With the increase of the field width, the oscillation of the energy spectra will become less obvious.
Third, for the time-dependent control fields, with the increase of the field frequency the oscillation of energy spectra becomes weaker and weaker, but the pair yield only has small changes.
When the field frequency reaches the same magnitude as $c^2$, the pair yield can be greatly improved because of the channel open of pair creation by multiphoton absorption.

Finally, we find that the information of the field direction, the oscillation frequency, and the time interval between two envelope fields encoded into the control fields can also be embodied in the change of the pair-creation rate with time. By decoding the pair-creation rate, the information encoded in the control fields can be received without using any matter as a propagation medium in the process of information transmission.

Our study further deepen the understanding of the effects of control fields on the energy spectra and the number of created pairs, and expand the scope of application of vacuum information transmission. These results can provide a significant reference to the related experiments in the future.

\begin{acknowledgments}
  The work is supported by the National Natural Science
  Foundation of China (NSFC) under Grants No. 11705278
  and No. 11974419, in part by the National Key R\&D
  Program of China under Grant No. 2018YFA0404802, and
  by the Fundamental Research Funds for the Central
  Universities (20226943).
\end{acknowledgments}

\appendix*
\section{Scattering of the Dirac equation in the double step potential}
There are two electric fields with the same direction and different magnitudes at $x=0$ and $x=-d$, see Fig. \ref{model}. The width of the electric field is ignored, so the space can be divided into three parts.
\begin{equation}
  V\left(x\right)=\left\{\begin{matrix}0\quad,&&x<-d&&( \mathrm{region \quad I})\\V_2\quad,&&-d\leq x\leq0&&( \mathrm{region \quad II})\\V_1+V_2\quad,&&x>0&&( \mathrm{region \quad III})\\\end{matrix}\right.
\end{equation}
Set $V_2 $ to be smaller than $2c^2$ , that is, the momentum in the second region is $k=\sqrt{\left(E-V_2\right)^2-c^4}/c$, then the wave function is set to
\begin{equation}
  \begin{matrix}
    \mathrm{\Psi}_{\rm I}=A\left(\begin{matrix}1\\\frac{pc}{E+c^2}\\\end{matrix}\right)e^{ipx}+A^\prime\left(\begin{matrix}1\\
    \frac{-pc}{E+c^2}\\\end{matrix}\right)e^{-ipx}\\
    \mathrm{\Psi}_{\rm II}=B\left(\begin{matrix}1\\\frac{kc}{E-V_1+c^2}\\\end{matrix}\right)e^{ikx}+B^\prime\left(\begin{matrix}1\\
    \frac{-kc}{E-V_1+c^2}\\\end{matrix}\right)e^{-ikx}\\
    \mathrm{\Psi}_{\rm III}=C\left(\begin{matrix}1\\\frac{-qc}{E_q-c^2}\\\end{matrix}\right)e^{iqx}+C^\prime\left(\begin{matrix}1\\
    \frac{qc}{E_q-c^2}\\\end{matrix}\right)e^{-iqx}\\
  \end{matrix}
\end{equation}
where
\begin{equation}
  \begin{matrix}
    E_q=V_1+V_2-E\\p=\sqrt{E^2-c^4}/c\\q=\sqrt{{-c}^4+\left(V_1+V_2-E\right)^2}/c\\
    k=\sqrt{{-c}^4{+\left(E-V_2\right)}^2}/c\\
  \end{matrix}
\end{equation}
At $x=-d$ and $x=0$, the wave function should be continuous, i.e.
\begin{equation}
  \begin{matrix}
    Ae^{-ipd}+A^\prime e^{ipd}=Be^{-ikd}+B^\prime e^{ikd}\\
    \gamma\left(Ae^{-ipd}-A^\prime e^{ipd}\right)=Be^{-ikd}-B^\prime e^{ikd}\\
    B+B^\prime=C+C^\prime\\
    \tau\left(B-B^\prime\right)=C-C^\prime\\
  \end{matrix} \label{a4}
\end{equation}
in the formula $\gamma=\frac{pc}{E+c^2}\frac{E-V_1+c^2}{kc}=\sqrt{\frac{E-c^2}{E+c^2}\frac{E-V_1+c^2}{E-V_1-c^2}}$ and
$\tau=\frac{kc}{E-V_1+c^2}\frac{V_1+V_2-E+c^2}{qc}\\=\sqrt{\frac{V_1+V_2-E+c^2}{V_1+V_2-E-c^2}\frac{E-V_1-c^2}{E-V_1+c^2}}$

Rewriting Eq. (\ref{a4}) into matrix form, it has the following form
\begin{equation}
  \begin{matrix}
    \left(\begin{matrix}A\\A^\prime\\\end{matrix}\right)=
  \frac{1}{2}\left(\begin{matrix}\frac{\gamma+1}{\gamma}e^{i\left(p-k\right)d}&
    \frac{\gamma-1}{\gamma}e^{i\left(p+k\right)d}\\\frac{\gamma-1}{\gamma}e^{-i\left(p+k\right)d}&
    \frac{\gamma+1}{\gamma}e^{i\left(k-p\right)d}\\\end{matrix}\right)\left(
      \begin{matrix}B\\B^\prime\\
      \end{matrix}
      \right)\\\left(
        \begin{matrix}
          B\\B^\prime\\
        \end{matrix}
        \right)=\frac{1}{2}\left(
          \begin{matrix}
            \frac{\tau+1}{\tau}&\frac{\tau-1}{\tau}\\\frac{\tau-1}{\tau}&\frac{\tau+1}{\tau}\\
          \end{matrix}
          \right)\left(
            \begin{matrix}
              C\\C^\prime\\
            \end{matrix}
            \right)\\
          \end{matrix}
\end{equation}
In the third region, there are no waves enter from the right hand side, so $C^\prime=0$
\begin{equation}
  \left(
    \begin{matrix}
      A\\A^\prime\\
    \end{matrix}
  \right)
  =\frac{1}{4}
  \left(
    \begin{matrix}
      \frac{\tau+1}{\tau}\frac{\gamma+1}{\gamma}e^{i\left(p-k\right)d}+\frac{\tau-1}{\tau}\frac{\gamma-1}{\gamma}e^{i\left(p+k\right)d}
      \\\frac{\tau+1}{\tau}\frac{\gamma-1}{\gamma}e^{-i\left(p+k\right)d}+\frac{\tau-1}{\tau}\frac{\gamma+1}{\gamma}e^{i\left(k-p\right)d}
    \\\end{matrix}
    \right)C
\end{equation}
The transmission coefficient is $T=\frac{C^\ast C}{A^\ast A}\frac{q}{p}\frac{E+c^2}{V_1+V_2-E}$
and finally, we get the expression
\begin{equation}
  T_E=\frac{4\gamma\tau}{{(\gamma\tau+1)}^2-(\gamma^2-1)(\tau^2-1){\sin}^2(kd)}
\end{equation}

%\bibliography{apssampNotes}

\end{document}